\documentclass[11pt]{article}
\usepackage{geometry}                

\usepackage{graphicx}
\usepackage{amssymb}
\usepackage{epstopdf}
\def\0#1{{\mathrm{#1}}}
\def\1#1{{\mathbb{#1}}}
\def\2#1{{\mathbf{#1}}}  
\def\3#1{{\mathcal {#1}}}
\def\4#1{{{\mathsf{#1}}}}     
\def\5#1{{{\widetilde{#1}}}}  
\def\6#1{{\overline{#1}}} 
\def\7#1{\breve{#1}}
 \def\8#1{{\widehat{#1}}}  
 \def\9#1{{\vec{#1}}}

\def\Alg{\mathop{{\0{Alg}}}\nolimits}
\def\alg{\mathop{{\0{alg}}}\nolimits}
\def\Bi{\mathop{{\0{Bi }}}\nolimits}
\def\apost{\mbox{\bf '}\,}

\def\miss{\mathrel{{\kern3pt\backslash\kern-8.1pt\bigcirc}}}

\def\Oplus{\bigoplus}
\def\O+{\bigoplus}
\def\LA+{\;\circ\hskip -9.1pt +\,}
\def\nogo{\mathrel{+\kern-9.0pt + \kern-12.4pt\to}}

 \def\supo{\mathaccent  "7017 }

\def\Bar{{{}\vrule height 8pt width 1pt depth 3pt{\kern3pt}}}
\def\<{{\left<\right.}}
\def\>{{\left.\right>}}

\def\Cliff{\mathop{{\0{Cliff}}}\nolimits}

\def\Grass{\mathop{{\V}}\nolimits}  
 
\def\Dim{\mathop{{\mathrm{Dim}}}\nolimits} 
\def\Diff{\mathop{{\mathrm {Diff}}}\nolimits} 
\def\diff{\mathop{{\4 {diff}}}\nolimits}

\def\Dual{\mathop{{\mathrm {Dual}}}\nolimits}

\def\hexp{\mathop{{\mathrm{hexp}}}\nolimits} 

\def\ph{\mathop{{\4{ph}}}\nolimits} 
\def\Qi{{\supo{\imath}}}

\def\so{\mathop{{\4 {so}}}\nolimits} 
\def\SO{\mathop{{\mathrm {SO}}}\nolimits} 
\def\slin{\mathop{{\4 {sl}}}\nolimits} 
\def\SL{\mathop{{\mathrm {SL}}}\nolimits} 
\def\Spin{\mathop{{\mathrm {Spin}}}\nolimits}
\def\spin{\mathop{{\4 {spin}}}\nolimits}
\def\su{\mathop{{\4 {su}}}\nolimits} 
\def\Texp{\mathop{{\mathrm {Texp}}}\nolimits} 

\def\Z{{\vrule width3pt height0pt depth0pt 
}}

\def\v{\vee}
\def\V{{\bigvee}}

\def\ox{\otimes}

\def\x{\times}

\def\BEQ{\begin{equation}}
\def\EEQ{\end{equation}}
\def\BEN{\begin{enumerate}}
\def\EEN{\end{enumerate}}
\def\BI{\begin{itemize}}
\def\BEI{\end{itemize}}
\def\BEA{\begin{eqnarray}}
\def\EEA{\end{eqnarray}}
\def\BTA{\begin{table}}
\def\ETA{\end{table}}

\def\fro{\leftarrow}

\def\cto{\kern3pt-\kern-9pt\succ}
\def\cfro{\prec \kern-9pt -\;}

\def\rar{\rightarrow}

\def\mapsfro{{\fro}\kern-4pt\rule{.5pt}{5pt}}

\def\dar{{\downarrow}}

\def\Diagr#1#2#3#4#5#6#7#8
{\begin{array}{rcccl}%
&              &{\scriptstyle #5}           &                                            &\cr
& #1                                          & \rar& #2                                      &\cr
\cr
{\scriptstyle #8}&\dar \;\; &         &\;\;\dar&{\scriptstyle#6} \cr
\cr
& #4                                   & \,\rar \,& #3                                      &\cr
&                           &{\scriptstyle#7}&  			             &     
            \end{array}}

\newtheorem{unitas}{}[section]

\def\unit{\unitas\rm}

\title{Quantum Simplicial Dynamics
} 
\author{David Finkelstein\\
{\normalsize Georgia Institute of Technology}}
\date{\today}                                      

\begin{document}

\maketitle

\abstract{
Present-day quantum field theory
can be regularized by a decomposition into quantum simplices.
This replaces the infinite-dimensional Hilbert space
by a high-dimensional spinor space
and singular canonical Lie groups by regular spin groups.
It radically changes the uncertainty principle 
for small distances.
Gaugeons, including the gravitational,
are represented as bound fermion-pairs, 
and 
space-time curvature as a singular organized limit
of quantum non-commutativity.

}

\section{Simplicial quantization}
 
\subsection{Quantization and de-radicalization}

One way to regularize
   the Standard Model and general relativity 
 while
preserving their agreement with empirical particle spectra and selection rules
 is to analysis them
into finite quantum simplicial elements.
This might be
a physical regularization, not a formal one.

A quantum theory represents its unmixed system input and outtake processes
by rays in a space of what Heisenberg called
``probability vectors".
A ``probability-amplitude vector" is more accurate;
a ``$\psi$ vector" or simply ``a $\psi$" is  quicker.
If the $\psi$ space is finite-dimensional, the operators of the theory
 have only finite spectra, and
 divergences  do not occur.
Such a quantum theory is called {\em regular} \cite{BOPP1950}.

Call a Lie algebra or group regular or singular 
according to whether its Cartan-Killing form is so.
Semisimple Lie algebras are regular;
 canonical Lie algebras are singular.

Quantum theories today use infinite-dimensional $\psi$'s 
for unitary representations of singular or
non-compact Lie algebras.

The singular Lie algebras include 
the canonical/Bose-Einstein Lie algebra 
\BEQ
\label{E:CCR}
\4h(M)=\alg(x^{\mu}, y_{\mu},i ): [x^{\mu}, y_{\mu'}]=\delta^{\mu'}_{\mu},\quad i\le \mu,\mu'\le M,
\EEQ
of  classical space-time coordinates and differentiators,
ordinary or gauge-covariant; of
quantum  coordinates and their canonical momenta in general;
 of the boson fields in particular;
and the functional algebras of gauge theory, including the $\diff$ of general covariance.
 
The non-compact Lie groups include the Lorentz group.
 Its Hilbert space representations are singular but
 its spinor representations are regular.
 
The radicals of the singular algebras indicate how singular they are.
Canonical quantization  reduced the radical dimensions of some Lie algebras of classical physics
from $\infty$  to  1  by  decontraction.
Here 
the remaining singular Lie algebras are
further reformed into simple subalgebras of $\spin(N,N)$ for some large finite $N$
determined by the number of quantum events in the system history.
This eliminates the radicals.

The limit $N\to\infty$ is not a contraction in the usual sense.
It goes by discrete
steps instead of a homotopy.
Yet in a weak sense still to be made precise,
 the step from $N$ to $N+1$ is small
when $N$ is large.
If $N$ is large enough, for example, 
the difference between systems of $N$ and $N+1$ quanta will be experimentally unresolvable
by present means.
Let us term a succession of small steps toward regularity a {\em reformation} 
whether they are discrete steps
or infinitesimal.

Quantizations have required ``painful renunciations", as Bohr put it.
Now
the ideas of space-time,  Hilbert space, and a fundamental theory
are renounced.
A typed and graded spinor space $\1S$ replaces  and elaborates the usual Hilbert space.
Its indefinite metric  has a familiar physical interpretation, 
essentially given by Dirac (\S\ref{S:Q}).

Classical systems with ancestrally finite sets for their sample spaces (= phase spaces)
are intrinsically finite in their properties (\S\ref{S:Q}) .
Finite sets are also simplices, whose vertices are their elements
and whose points are the convex statistical mixtures of their vertices.
Quantum simplices replace the classical statistical mixture by quantum superposition  (\S\ref{S:Q}).

Vertices of a simplex may themselves be simplices of a deeper level.
Call a simplex   a {\em complex} when we wish
 to emphasize that it is not only finite but ancestrally finite,
 finite
 all the way down.

The quantum complex of
\S\ref{S:Q}
is regular, with its $\psi$'s
in  a spinor space,  finite dimensional by definition,
and seems 
rich enough to represent any finite quantum structure
and any of the experimental transformation groups of quantum physics
as closely as necessary.
It seems closer to experiment to use $\psi$'s in a spinor
than in a Hilbert space.

From a logic of yes-or-no questions about one individual
one can construct a logic of how-many questions about a quantity of such individuals.
Such  a construction  is called
{\em quantification}
after William Hamilton, ca. 1850.
The higher-order algebra of sets (or classes) is a quantification theory.
It transforms a theory with sample space $S$ into a quantified theory with sample space $2^S$,
the power set of $S$.

As von Neumann pointed out,
when Heisenberg replaced
the commutative algebra of dynamical variables
by his non-commutative  algebra of dynamical variables,
he effectively revised the first-order logic of physical systems;
for  predicates are merely two-valued dynamical variables.
Quantum theory replaced Boolean logic
by a projective logic.

This destroyed  classical quantification
theory at its Boolean base and necessitated a quantum 
 replacement.
Several quantum quantifications were swiftly constructed;
but from scratch, 
as if classical quantification had never existed.
Classical  quantifications include such assemblies as
combinations and permutations, sets and sequences.
Quantum sets were instead called  Fermi-Dirac ensembles.
Quantum sequences were called Maxwell-Boltzmann ensembles.
These quantifications convert each one-quantum $\psi$ 
into an annihilation/creation (or A/C) operator $\8\psi$, 
 acting on many-quantum 
$\psi$ vectors,
and generating an algebra that determines the statistics.

Call a classical sequence 
$(e,e,\dots, e)$ of any number of equal elements,
a {\em sib}.
This construct was too useless to merit a name
before the advent of the quantum
theory. 
A quantum sib, however,
is a Bose-Einstein ensemble,
whose every  $\psi$ is a superposition of  tensor powers like $\9e\ox\dots\ox\9e$,
where $\;\9{}\;$ indicates a  $\psi$ vector.

All the classical modes of quantification mentioned
 are easily expressed in terms of power sets, but
the converse is not true.
Moreover, the only regular quantification theory among thm
is the quantum set, the Fermi-Dirac ensemble,
whose $\psi$ space is a Grassmann algebra.
This suggests that if we choose to start with just one kind of quantification,
even as a toy theory,
we ought to try the Fermi-Dirac kind first.
Let us do so.

\subsection{The spinor space $\1S$}
\label{S:Q}

Since space-time 
belongs to a deeper logical type
than the fields that are functions on it, 
their joint quantization 
calls for a quantum theory of several levels.

{\em  Unitization} (Glaserfeld's term) is the bracing operation
\BEQ
\iota: a\mapsto \{a\}\equiv \iota a \equiv\6a\/.
\EEQ
Unlimited
iteration
 of the classical operations $1$, $\iota$ and $\v$ generates
  an infinite set $\8{\1S}$ of classical (finite) complexes.
  
Correspondingly,
the $\psi$ space $\1S$
of the generic quantum complex is finitely generated
by  $+$ and $\1R$
for  quantum superposition, 
the unitization operator or unitizor $\iota$, 
subject to linearity,
and the Grassmann product $\vee$.
 
Iterated unitization
 is standard in classical thought;
Peano used it to generate the natural numbers.
It already occurs to a limited extent at several junctures in quantum physics today;
as when we treat helium nuclei as point particles.
There $\iota$ is a phenomenological description
 of a more complex process
 of dynamical binding of four fermions.
Likely $\iota$ is phenomenological here too.

 Write the set  of all finite subsets of a set $s$ 
 as  $S=\V s$ or $2^s$.
 Let us call the set $s$ the {\em logarithm} of $2^s$.
For example, $\8{\1S}$ is its own logarithm: $\8{\1S}=2^{\8{\1S}}$.

.
As usual,  use the same symbol for 
corresponding classical and quantum constructs
when context makes it clear which is meant.
When necessary, 
designate a quantum correspondent of $x$ by $\9x$.
For example, write  the Grassmann or exterior algebra
 of a vector space $\3V$ as $\V \3V\equiv \22^{\3V}$.
This is the vector space spanned by  Grassmann products 
of finitely many vectors in $\3V$.

Like $\8{\1S}$, $\1S$ is its own logarithm,
$\1S=\V \1S = 2^{\1S}$,
and is doubly graded, by Grassmann grade $g$ and Quine type $T$:
\BEQ
\1S=\Oplus_{g,T}\1S^{(g)}[T]
\EEQ
The spinors of $\1S$ are  used as $\psi$'s for quantum complexes.

These spinors carry many physical variables besides spin.
For clarity, call a quantum complex a  {\em plexon}, and call a $\psi$ vector for 
a plexon a {\em plexor}.
Spinors are special cases.

Since simplicial quantum theory replaces 
the Hilbert space of canonical quantum theory by a Grassmann algebra $\1S[E]$, 
it is a graded (or ``super-'') 
quantum theory of an extreme kind.
In the  terms of Bryce DeWitt, plexons are
all soul and no body.

In nature, all elementary fermions seem to have spin $1/2$.
The systems represented in the Grassmann algebra $\1S$
are fermions.
Therefore give each type $\1S$ 
the structure of a spinor space,
not a Hilbert space  (\S\ref{S:PAULI}).
The Cartan theory of spinors
enables us to assign
 a spinor structure to every Grassmann algebra $\1S[T]$.
The underlying quadratic space is the direct sum
\BEQ
\3W[T]:=\Bi \1S^{(1)}[T]:=\1S^{(1)}[T]\oplus \Dual \1S^{(1)}[T].
\EEQ
The quadratic space underlying the spinor space $\1S$
is the bipolar space 
of (\ref{E:BI}).

In homological algebra, a  {\em complex}---a collection of simplices--- is 
usually represented
as a sum of simplices.
But
in $\1S$,  addition is mere quantum superposition, not collection.
It does not increase  the number of simplices
 but merely changes the possibility for one simplex.
One assembles plexons into a plexon of higher type
according to quantum logic;
not by adding them
 but by multiplying their unitizations.
The vertices of the resulting complex are the unitizations of the  simplices of the type below.
A complex is a simplex of simplices, not a sum of simplices.

\subsection{Index conventions}
 \rule{0pt}{0pt}
 \vskip-10pt
 $\1S[T]:=$ the subspace of $\1S$ consisting of all polyadics of type $T$.
 
 $\Bi \1S[T]: =\1S^{(1)}[T]\oplus \Dual \1S^{(1)}[T]$.

$1_{t}:=$ a typical polyadic basis spinor in the Grassmann algebra $\1S[T]$, 
indexed  by  a lower-case version
of $T$\/, the symbol for the type.

$1_{\6 t}:=$  a typical monadic basis vector in the first-grade subspace $\1S^{(1)}[T+1]$:
\BEQ
1_{\6 t} :=\6{1_t} :=\iota (1_t)=:1_{(2^t)}\/.
\EEQ

$1_{\ddot t}:=$
a typical   basis vector for the bispace $\Bi \1S[T+1]$.

In the classical basis $t$ is the serial number of $1_t$.
$t$ and 
$\6t$  take on  $\hexp T$ values.
$\ddot t$ takes on $2\hexp T$ values.

\subsection{Bar codes}
\label{S:BAR}

An algebraic {\em bar code} is easier to read and write than
a graph of a multidimensional complex.
Construct a bar code  from a formula for the complex
 in the symbols $1$, $\iota$, and $\v$,
by omitting the symbols 1 and $\vee$ as usual, 
 writing $\iota x$ as the bar symbol $\6 x$,
 ordering  factors by their serial numbers,
and write superpositions as usual.
$\1S$ is the vector space of quantum bar codes
(Table 1).
{\Huge
\BTA [ht]
\label{T:BARCODES}
\caption{Bar codes by rank $r$  and serial number $n$}
{\boldmath
\[
\label{E:BARCODES}
\begin{tabular}{|c|ccccccccccccc|} 
 \hline
{ 6} \vrule height 35pt depth 0pt width 0pt 
&
\huge$\stackrel{\6{\6{\6{\6{\6{\6{\Z}}}}}}}{}$&\huge$
\stackrel {\6{\6{\6{\6{\6{\6{\Z}}}}}}\,\6{\Z}}{}$&\huge$
\stackrel {\6{\6{\6{\6{\6{\6{\Z}}}}}}\,\6{\6{\Z}}}{}$&\huge$
 \stackrel {\6{\6{\6{\6{\6{\6{\Z}}}}}}\,\6{\6{\Z}}\,\6{\Z}}{}$&\huge$
\stackrel {\6{\6{\6{\6{\6{\6{\Z}}}}}}\,\6{\6{\6{\Z}}}}{}$&\huge$
\stackrel {\6{\6{\6{\6{\6{\6{\Z}}}}}}\,\6{\6{\6{\Z}}}\,\6{\Z}}{}$&\huge$
\stackrel {\6{\6{\6{\6{\6{\6{\Z}}}}}}\,\6{\6{\6{\Z}}}\,\6{\6{\Z}}}{}$&\huge$
\stackrel {\6{\6{\6{\6{\6{\6{\Z}}}}}}\,\6{\6{\6{\Z}}}\,\6{\6{\Z}}\,\6{\Z}}{}$&\huge$
\stackrel {\6{\6{\6{\6{\6{\6{\Z}}}}}}\,\6{\6{\6{\Z}}\,\6{\Z}}}{}$&\huge$
\stackrel {\6{\6{\6{\6{\6{\6{\Z}}}}}}\,\6{\6{\6{\Z}}\,\6{\Z}}\,\6{\Z}}{}$&\huge$
\stackrel{\6{\6{\6{\6{\6{\6{\Z}}}}}}\,\6{\6{\6{\Z}}\,\6{\Z}}\,\6{\6{\Z}}}{}$&\huge$
\stackrel{\6{\6{\6{\6{\6{\6{\Z}}}}}}\,\6{\6{\6{\Z}}\,\6{\Z}}\,\6{\6{\Z}}\,\6{\Z}}{}$&$
 ^{\dots}
 $
 \cr
  \vrule height 12pt depth 0pt width 0pt&$\!\!^{\hexp 6}$&$^{\dots}$&&&&&&&&&&&\cr
\hline
{ 5}&
\vrule height 30pt depth 0pt width 0pt
\huge$
\stackrel {\6{\6{\6{\6{\6{\Z}}}}}}{}$&\huge$
\stackrel{\6{\6{\6{\6{\6{\Z}}}}}\,\6{\Z}}{}$&\huge$
\stackrel{\6{\6{\6{\6{\6{\Z}}}}}\,\6{\6{\Z}}}{}$&\huge$
\stackrel{\6{\6{\6{\6{\6{\Z}}}}}\,\6{\6{\Z}}\,\6{\Z}}{}$&\huge$
\stackrel{\6{\6{\6{\6{\6{\Z}}}}}\,\6{\6{\6{\Z}}}}{}$&\huge$
\stackrel{\6{\6{\6{\6{\6{\Z}}}}}\,\6{\6{\6{\Z}}}\,\6{\Z}}{}$&\huge$
\stackrel{\6{\6{\6{\6{\6{\Z}}}}}\,\6{\6{\6{\Z}}}\,\6{\6{\Z}}}{}$&\huge$
\stackrel {\6{\6{\6{\6{\6{\Z}}}}}\,\6{\6{\6{\Z}}}\,\6{\6{\Z}}\,\6{\Z}}{}$&\huge$
\stackrel {\6{\6{\6{\6{\6{\Z}}}}}\,\6{\6{\6{\Z}}\,\6{\Z}}}{}$&\huge$
\stackrel{\6{\6{\6{\6{\6{\Z}}}}}\,\6{\6{\6{\Z}}\,\6{\Z}}\,\6{\Z}}{}$&\huge$
\stackrel{\6{\6{\6{\6{\6{\Z}}}}}\,\6{\6{\6{\Z}}\,\6{\Z}}\,\6{\6{\Z}}}{}$&\huge$
\stackrel{\6{\6{\6{\6{\6{\Z}}}}}\,\6{\6{\6{\Z}}\,\6{\Z}}\,\6{\6{\Z}}\,\6{\Z}}{}$&$
^{\dots}
 $\cr
& $\!\!^{\hexp{5}}$&$^{\dots}$&$$&$ {}$&&&&&&&&&\cr
\hline
{ 4}& \vrule height 25pt depth 0pt width 0pt
\huge $
\stackrel {\6{\6{\6{\6{\Z}}}}}{}$&\huge$   
\stackrel{\6{\6{\6{\6{\Z}}}}\,{\6{\Z}}}{}$&\huge$
\stackrel{\6{\6{\6{\6{\Z}}}}\,{\6{\6{\Z}}}}{}$&\huge$
\stackrel{\6{\6{\6{\6{\Z}}}}\,{\6{\6{\Z}}\,\6{\Z}}}{}$&\huge$
\stackrel{\6{\6{\6{\6{\Z}}}}\,{\6{\6{\6{\Z}}}}}{}$&\huge$
\stackrel{\6{\6{\6{\6{\Z}}}}\,{\6{\6{\6{\Z}}}}\,\6{\Z}}{}$&\huge$
\stackrel{\6{\6{\6{\6{\Z}}}}\,{\6{\6{\6{\Z}}}\,\6{\6{\Z}}}}{}$&\huge$
\stackrel {\,\6{\6{\6{\6{\Z}}}}\,{\6{\6{\6{\Z}}}\,\6{\6{\Z}}}\,\6{\Z}}{}$&\huge$
\stackrel {\6{\6{\6{\6{\Z}}}}\,{\6{\6{\6{\Z}}\,\6{\Z}}}}{}$&\huge$
\stackrel {\6{\6{\6{\6{\Z}}}}\,{\6{\6{\6{\Z}}\,\6{\Z}}}\,\6{\Z}}{}$&\huge$
\stackrel {\6{\6{\6{\6{\Z}}}}\,{\6{\6{\6{\Z}}\,\6{\Z}}}\,\6{\6{\Z}}}{}$&\huge$
\stackrel {\6{\6{\6{\6{\Z}}}}\,{\6{\6{\6{\Z}}\,\6{\Z}}}\,\6{\6{\Z}}\,\6{\Z}}{}$&
$^{\dots}$\cr
 &16 &17&18&19&20&21&22&23&24&25&26&27&$^{\dots}$\cr
\hline 
{ 3} & \vrule height 20pt depth 0pt width 0pt 
\huge$
\stackrel {\6{\6{\6{\Z}}}}{}$&\huge$  
\stackrel{{\6{\6{\6{\Z}}}\,\6{\Z}}}{}$&\huge$ 
\stackrel{\6{\6{\6{\Z}}}\,\6{\6{\Z}}}{}$&\huge$
\stackrel{{ \6{\6{\6{\Z}}} \, \6{\6{\Z}} \, \6{\Z}}}{}$&\huge$
\stackrel{{\6{\6{\6{\Z}}\,\6{\Z}}}}{}$&\huge$
\stackrel{{\6{\6{\6{\Z}}\,\6{\Z}}}\,{\6{\Z}}}{}$&\huge$ 
\stackrel{{\6{\6{\6{\Z}}\,\6{\Z}}}\,{\6{\6{\Z}}}}{}$&\huge$
\stackrel{{\6{\6{\6{\Z}}\,\6{\Z}}}\,{\6{\6{\Z}}}\,\6{\Z}}{}$&\huge$
\stackrel{{\6{\6{\6{\Z}}\,\6{\Z}}}\,{\6{\6{\6{\Z}}}}}{}$&\huge$
\stackrel{{\6{\6{\6{\Z}}\,\6{\Z}}}\,{\6{\6{\6{\Z}}}}\,\6{\Z}}{}$&\huge$
\stackrel {{\6{\6{\6{\Z}}\,\6{\Z}}}\,{\6{\6{\6{\Z}}}}\,\6{\6{\Z}}}{}$&\huge$ 
\stackrel{{\6{\6{\6{\Z}}\,\6{\Z}}}\,{\6{\6{\6{\Z}}}}\,\6{\6{\Z}}\,\6{\Z}}{}$&\cr
 &4 &5&6&7&8&9&10&11&12&13&14&15&\cr
\hline
{ 2}  & 
\huge
 $
\stackrel {\6{\6{\Z}}}{}$&\huge$
\stackrel{{\6{\6{\Z}}}\,\6{\Z}}{}$&&&&&&&&&&&\cr
&2&3&&&&&&&&&&&\cr
 \hline
{ 1} & 
\Huge $\stackrel {\6{\Z}}{}$&${}$&&&&&&&&&&&\cr
&1 &&&&&&&&&&&&\cr
\hline
{ 0}  &
 &&&&&&&&&&&&\cr
&\kern3pt  0 &&&&&&&&&&&&\cr
\hline
\hline
{\em r}  &\vrule height 15pt depth 0pt width 0pt&&&&&&{${1_n}$}&&&&&&\cr
&&&&&&&{ \em n}\vrule height 16pt depth 0pt width 0pt&&&&&&\cr
\hline
\end{tabular}
\]
}
\ETA
}

Each bar code describes an oriented simplicial complex,
whose vertices are its factors,
oriented by the order of their multiplication.
They must be read from the bottom up.
Which bars stand for the same vertex 
is determined by what lies below them and not by what lies above them.
The bar codes
constructed without
quantum superposition, using only $\vee$ and $\iota$,
 form a classical basis of $\1S$.
Their signs are fixed by serially ordering their monadic factors.

A $g$-adic is a simplex of  exactly $g$  vertices.
Each vertex of a simplex  is a segment of its bar code covered by a single bar,
as long as necessary, representing a monadic. 

The rank of a polyadic is the height
in bars of its highest monadic factor.

A simplicial quantum theory of type $T$ is one whose plexors
have  type $\1S[T]\subset \1S$. 

Assume that there exist:
\begin{enumerate}
\item a cellular plexon type $\1S[C]$
that supports
the Lie algebra of 
 spin and the unitary charges.
\item
an event type $\1S[E]$ that supports the orbital variables as well.
\item a field type $\1S[F]=\1S[E+1]$ 
that supports field variables as well.
\end{enumerate}
Let us roughly
estimate the event type   $\1S[E]$ 
whose dimension $N$  is large enough 
to pass for the Hilbert space of a fermion.
Provisionally,  $C=3$, $\Dim \1S[C]= \hexp 2=16$
might suffice for the spin and charge of one cell.
Then $E=4$  would imply $N \le  \hexp 4={(2^{16})}$, far too small for a 
quasi-continuum of events;
while $E=5$ implies $N\le \hexp 5=2^{64\0K}$  (where $\0K:=2^{10}$),
far more than required.

The Poincar\'e group can be approximated within present experimental error,
 though non-uniformly,
by a subgroup of  $\SO(\1S[5])$ but not $\SO(\1S[4])$.
The  Standard Model groups seem to be faithfully represented
 in $\1S[5]$ up to experimental accuracy.
 
Therefore assume tentatively that $C=3$,  $E=5$,
and   $F=6$.

\section{Space-time quantizations}

\subsection{Yang quantum space}

Most physical theories today assume an absolute space-time,
with many reference frames related by a relativity group.
To simplify  the Poincare-Heisenberg Lie algebra 
\BEQ
\ph(4)=\alg(x^m,p_m,L_{m'm},i)
\EEQ
 of special relativistic mechanics,
Yang dropped this assumption and adopted the decontraction
$\ph(4) \cfro \so(5,1)$ \cite{YANG1947}.
This relativizes the split into position space and momentum space,
 and quantizes the imaginary $i$ of the commutation relations.
He retained a complex Hilbert space representation of infinity dimensions, however,
with its separate central $i$.

Many  reformations decontract the canonical Lie algebra $\4h(1)$
to the sphere Lie algebra $\so(3)$:
\BEQ
\label{E:ROTATOR}
\so(3):\quad [p,q]=r, \quad [q,r] = \lambda p, \quad   [r,p]=\lambda q,\quad \lambda \to 0,
\EEQ
replacing  the canonical $i$ by a variable $r$, 
and the singular Killing form by a regular one with negative-definite Killing form.
 Slightly different reformations de-contract $\4h(1)$
  to the hyperboloid Lie algebra:
 \BEQ
\label{E:HROTATOR}
\so(2,1):\quad [p,q]=r, \quad [q,r] = \lambda p, \quad   [r,p]=-\lambda q, 
\quad \lambda \to 0,
\EEQ
with indefinite Killing form.
Another decontraction  of
the canonical Lie algebra  $\4h(1):=\alg(q,p,i)$ of (\ref{E:CCR})
acts on a spinor space instead of a Hilbert space,
representing  $\so(3)$ by dyadic spin operators in a Clifford algebra
$\1S[2]\ox \Dual \1S[2]$, of type 2:
\BEQ
p=\gamma_{23}, \quad  q=\lambda \gamma_{31}, \quad r=\lambda \gamma_{12}, \quad \lambda\to 0.
\EEQ
A unification of position and momentum occurs 
 in other gauge theories in quantum space-time
\cite{MADORE2000}\/.

\subsection{Feynman quantum space}

Feynman proposed a spinorial space-time quantization that in retrospect
fits into the neutral spinor space $\1S$, not a Hilbert space $\3H$.
He wrote the coordinates as sums of many commuting Dirac matrix-vectors:
\BEQ
\label{E:FEYNMAN}
x^m\sim\gamma^m(1)\oplus \dots\oplus \gamma^m(N)\/.
\EEQ
This was undoubtedly suggested by the relativistic proper-time equation of motion
\BEQ
\frac{dx^m}{d\tau}=\gamma^m
\EEQ
for a Dirac particle.
These Dirac spin operators $\gamma^m(n)$ 
belong to a quantum element of space-time.
(\ref{E:FEYNMAN}) concerns an assembly
of $\4N$ such elements with  a Palev statistics,
which had not been discovered yet.
An implied quantum unit of time $\4X$ 
is omitted from (\ref{E:FEYNMAN}).

Feynman's proposal  must be extended to represent momentum and $i$.
This can be done
within the same Dirac algebra,
as  noted by Marks \cite{MARKS2008}.
For  the Standard Model fermions, $i\gamma^{4321}$ is $\pm 1$.
This suggests that at a deeper level $\gamma^{4321}$ performs some of the functions of $i$.
If we set
\BEA
\label{E:MARKS}
p_m&\sim&\gamma_{4321}(1)\gamma_m(1)\oplus \dots\oplus \gamma_{4321}(N)\gamma_4(N),
\cr
i&\sim& \frac{\gamma_{4321}(1)\oplus \dots\oplus \gamma_{4321}(N)}N,
\EEA
then the commutator of $\delta x$ and $\delta p$ has a $\gamma^{4321}$ where the canonical theory has $i$.
(\ref{E:MARKS}) omits a quantum unit of energy
\BEQ
\4P=\frac{\hbar}{\4N\4X},
\EEQ
required to balance the units.
 (\ref{E:MARKS}) defines a simplicial quantum space.
In the limit of classical space-time, 
$
\4X\to 0$, $\4P\to 0$, and $\4N\to \infty$.
The spin  now serves as an element of space-time
and momentum-energy, as well as 
angular momentum $\delta L_{m'm}=\gamma_{m'm}/2$.
Soon its simplex will be enlarged to carry the unitary charges as well, and 
to describe entire events.
When it is necessary in order to reduce confusion, 
call a spin with this enlarged physical interpretation and algebra,
a  quantum simplex, or {\em plexon}\/.

The space-time coordinates, 
Lorentz angular momenta, momenta, and $i$ of a plexon of the cellular type
correspond 1-1 to $4+6+4+1=15$ traceless $4\x 4$ 
Dirac matrices in the real Dirac spinor space $4\1R$
of the plexon, soon to be enlarged.
The commutator Lie algebra of $\Cliff(3\1R\ominus 1\1R)$
is $\slin(4)\cong \so(3,3)$ instead of the Yang $\so(5,1)$.
The Feynman quantum space has
an automorphism  $\gamma^m\mapsto\gamma^{4321}\gamma^m$
of the Clifford algebra of polynomials in the $\gamma^m$;
it exchanges space-time with momentum-energy.
  
The Feynman quantum space is not represented in Hilbert space here
but  in a spinor space $\1S[E]$, and
 so it  is regular.
If the extra two coordinates of $\so(3,3)$ are frozen out
in the condensation that produces $i$,
as assumed,
their peculiar timelike signature seems to lead to no 
contradictions with macroscopic experience.

Each type $\1S[T]$ is  a spinor space
for a quadratic vector space $\3W[T]$.
The vector-wise unitization $\iota\apost \1S[T]=\1S^{(1)}[T+1]$
is
the  null semivector space for the next spinor space, of type $T+1$, 
depending on context.
This context-dependence can be eliminated merely by explicitly tagging
spinors and semivectors differently,
so it seems harmless.

As spinor space,
the spin group $\Spin[T]$ doubly covers the orthogonal group $\SO[T]$
of the 
{\em bipolar space} 
\BEQ
\label{E:BI}
\3W[T]:= \Dual \1S^{(1)}[T]\oplus \1S^{(1)[T]}=:\Bi \1S^{(1)}[T]
\EEQ
with neutral {\em bipolar norm} $b: \3W\to \Dual \3W$ given by
\BEA
\forall Q\in \1S, \; \forall Q'\in \Dual \1S,\; W&:=&Q+Q': \cr
  bW\circ W:=\|W\|=\|Q+Q'\|&:=& Q'\circ Q.
\EEA
The bipolar space is also called the quantum space \cite{SALLER2006a}.
A basis $I_s$ for monadics in $\1S$ defines a basis $I_{\ddot s}$ for the
bipolar vectors of  $\3W$.
These in turn define a basis $\gamma_{\ddot s}$ for the first grade of 
the Clifford algebra over $\3W$.
Grassmann left-multiplication
defines  the first grade generator $\gamma_{s}$ of the type-$T$ Clifford algebra 
$\Cliff  \3W^{(1)}[T]$:
\BEQ
\forall \psi\in \1S^[T], I_{ s}\in \3W[T-1] : \gamma_{s} \psi:=I_{s}\vee \psi\/.
\EEQ
And the dual basis elements $\gamma^{s}$  are dually defined by
Grassmann left-differentiation with respect to $I_s$.

The Dirac spin  operators $\gamma^m$ come from  a low type $C$ of this Clifford algebra.
The Fermi-Dirac annihilation/creation operators $\psi(x)$ belong to a high type $E$ of the same Clifford algebra.

The indefinite metric $b$ of $\3W$ gives rise to that of special relativity,
which distinguishes the forbidden spacelike directions
from the allowed timelike ones.

\subsection{Neutral spinor form}
\label{S:PAULI}

Any spinor space
has a 
bilinear {\em spinor form}\/,
  that is
invariant under its spin group $\Spin[T]$.
Cartan called it $C$; 
in electron theory it is often designated by $\beta$.
For an irreducible spinor space, $\beta$ is unique up to a 
numerical factor.

The addition of a quantum $A$ to a system is equivalent to the
removal of a quantum $B=\Dual A$ from the system.
The anti-particle $\6A$ of a particle  $A$
with energy  $E$ is a dual particle with energy $-E$.
 
Consider a basis $\gamma^{m\pm}$ of Clifford algebra generators
that are anti-commuting square roots of $\pm 1$.
Let $\Pi_{\pm}$ be the product of  the operators 
$\gamma^{m\pm}$ that have square  (say) $\pm 1$. 
Let $1_s$ be  a spinor basis in which the $\gamma_{m+}$ are symmetric
and the $\gamma_-$ are skew-symmetric.
Let $E=\sum 1_s\ox 1_s$ be the Euclidean quadratic form that is 
diagonal in the basis $1_s$.
Then two spinor forms, possibly proportional to each other, are
\BEQ
\beta_\pm =\Pi_ {\pm} E
\EEQ
Write $\beta[T]$ for the spinor form of $\1S[T]$.
Then $\beta[2]$ is skew-symmetric and
$\beta[T]$ is neutral and symmetric for $T>2$.

The physical interpretation of the indefinite metric 
 is borrowed from Dirac's electron theory:
 The neutral spinor metric $\beta$ distinguishes input operations from outtake
operations  
(\S\ref{S:BIPOLAR}).

The normed space  $\3W[2]=\1S[2]\oplus \Dual \1S[2]$ of type 2
is a neutral quadratic space of eight dimensions.
It completes the first cycle of the Bott periodicity.
Its reduced spinor and dual-spinor spaces are also of eight dimensions,
and are related to $\3W[2]$ by triality.

One may choose any type $C$ of $\1S$ to be the $\psi$ space for an ancestral cell,
and analyze any spinor of the event type $E>C$
as a complex of clones of the cell $C$.
Then any transformation $L: \1S[C]\to \1S[C]$  induces a transformation
$\sum_C^E  L: \1S[T]\to \1S[T]$ for all $T > C$,
the {\em cumulant} of $L$.

If $X$ is any basic polyadic in $\1S[E]$, a product of basic monads,
then by masking all bars above those of type $C$
one exhibits a collection of polyads of type $C$,
whose transforms under $L$ are well defined.
To transform the complex $X$, one transforms  each of these
type-$C$ polyads in $X$ in turn and
unmasks.
The sum of all these transforms is $\sum_C^{C+1} L$.
This is extended to general polyadics by linearity.

In this way any representation of a Lie algebra on the ancestral cell extends to one
 on the entire complex,  its cumulant.
 $\1S$ is a set theory with null foundation.
 This construction re-interprets it  as a set theory with foundation represented by $\1S[C]$.
 
 For example, the cell space $\1S[2]$ can be identified with real Dirac spinor space.
Then all higher-type spinors of $\1S[T]$, $T>2$,
 represent simplices composed of spins 1/2
by iterated Fermi quantification.
The Dirac spin vector $\gamma^m$ for the generic fermion simplex is part of 
the spin vector $\gamma^c$ 
of its ancestral  cell.
The total spin angular momentum $J_{c'c}$ is the cumulant over the system complex,
of the cell spin
angular momenta $\gamma_{c'c}$ of its constituent cell simplices.

This builds in the spin-statistics correlation as an identity.

The orthogonal group $\SO(4,4)$ of $\3W[2]$
does not include 
Yang  $\SO(5,1)$ but it includes the simplicial quantum $\slin(4)=\spin(3,3)$,
which in turn includes Lorentz $\spin(3,1)$.

The two extra  timelike dimensions of $\spin(3,3)$ are supposed to freeze out in the 
condensation of $i$.

Classical space-time is supposed to emerge as a statistically smoothed 
approximation to an organized form of such a quantum cellular 
complex.

Isospin and color, the unitary charges,
 must commute with spin and orbital angular momentum.  
They may be represented  by operators 
on the extra 12 spinor dimensions
of the ancestral spin cell.

\subsection{Quantization as quantification}
\label{S:QQ}

Let the generic one-quantum $\psi$ be expanded in a monadic basis $\{1_q\}$
with numerical amplitudes $\psi^{q}$:
\BEQ
\psi=\sum_{q}\psi^{q} 1_q=\psi^{q}1_{q}\/.
\EEQ
``Second quantization" replaces the numerical coefficients $\psi^q$ by operators $\8\psi^q$.
This is often said to misconstrue what is not a quantization at all,
since it introduces no quantum constant or homotopy parameter,
but a mere quantification,  going from one quantum to many,
replacing the $\psi$ vector  $1_q$, 
not its numerical coefficient, by an annihilation/creation operator $\81_q$.

But the difference between $\8\psi^q$ and  $\81_q$ is a mere duality.
This suggests deeper connections between quantization and quantification
going both ways:

1. Quantum quantification is also a second quantization in a generalized sense. 
In a certain singular organized limit, in which the number of quanta $N\to \infty$,
a quantum field theory becomes a classical field theory.
Were the quantum field theory a true quantization,
the return to a classical field theory would be a contraction, a homotopy.
In fact the limit $N\to \infty$  can be regarded as a generalized contraction,
one that proceeds  by small discrete steps instead of by a homotopy.
In this extended sense, a quantification is also a quantization.
The small quantum constant that it introduces is $1/N$.
In an organized limit $1/N\to 0$ the field operators are centralized.

2.  Canonical quantization entails a quantification with Bose statistics.
To see this,  break canonical quantization down into two steps:
\begin{enumerate}
\item {\em Atomize}\/.  Select and decontract a Lie algebra $a$ of operators, 
to become the variables of the quantum atomic element. 
\item {\em Quantify}\/.  Form the variables of a quantity of quantum atoms
as  the polynomials in $a$ and limits thereof.
\end{enumerate}
The elements of $\8a$ represent operations that  put in and take out  quanta
of excitation,
 such as phonons.
For the harmonic oscillator, for example,
$a$ is spanned by the elements $q$ and $p$, 
and is the $\psi$ space of one phonon.
The polynomials in $p$ and $q$, and their limits,
correspond to symmetric tensors over $a$.
The quantized harmonic oscillator is then a bosonic assembly of phonons,
a quantified phonon.

The point is that a regular quantum theory requires a quantization 
based on a regular statistics like the Fermi or Palev statistics.
 Bose statistics does not work.

The corresponding steps of $\1S$ simplicial quantization are:
\begin{enumerate}
\item {\em Atomize}\/.  
Select a graded Lie algebra $a$ of basic system variables
 and decontract it into  a subalgebra $\8a$ of
$\1S[E]\subset \1S$ of simplicial quantum events.
\item {\em Quantify}\/.  Form the  Grassmann algebra in $\1S[E]$ generated by $\8a$.
\end{enumerate}
This is a Fermi quantification.

 The simplicial quantum theory will use at least six successive quantifications.
During the  contraction of a simplicial quantum theory
 to a canonical classical or quantum theory, moreover, 
 the type $T$ increases beyond any finite bound.
 
The Grassmann functor $\Grass$
quantifies, converting a
 one-quantum $\psi$ space to a many-quantum one, and also gauges,
cloning one cell Lie algebra into many isomorphs.
Any coordinate system  $(x^1,x^2, x^3, x^4)$ 
orders space-time events into a sequence of sequences of sequences of sequences,
implying that events have Maxwell-Boltzmann statistics.
No such quanta are found in nature.
Fermions and bosons are found, and bosons are singular,
so let us assemble space-time as a
quantum simplex of simplices of $\dots$  of simplices.

Cartan constructed
spinors as elements of the exterior algebra 
over a null ``semivector" space.
Chevalley noted that Grassmann algebras describe simplicial complexes.
Combining their insights,
 each quantum simplicial cell with $n$ vertices, 
is described by a spinor,
which we have called a plexor to avoid spacial connotations,
and is therefore a spin, or plexon.

The plexor has 
$2^n$ components, one for each possible simplicial face.
Further structure resides in the deeper levels, covered by one or more bars, 
successive unitizations.
In the singular continuum limit, $\4X\to 0$\/, $\4N\to \infty$.

In $\1S$, Grassmann grade counts simplicial quantum events, 
Quine type counts nested $\iota$'s,  
and the basic type-$E$ dynamical operators $J^{A}[E]\in \so[E]$,
 being  angular momentum operators
 of an orthogonal group $\SO[E]$,
count  angular momentum in units of the roots of the Lie algebra 
$\so[E]$\/.
$\so[C]$ respects the neutral quadratic form $\4b[C-1]$,
which contracts to the Minkowski metrical form.

The action of $\so(3,3)$ on type $C+1$
is induced by its action on the ancestral cell of type $C$.
Let us call this familiar summation process and its iterates {\em cumulation}, and write it as
$J[C+1,C]=\Sigma J[C]$.
The action on type $E$ is then the cumulant $J[E,C]=\Sigma^{E-C} J[C]$.

\section{Simplicial quantum events} 

\label{S:QSPACETIME}

At first Einstein described a space-time point or  event operationally
as a smallest possible occurrence, like
the collision of two small hard bodies.
It is oddly anachronistic to take
collisions of classical macroscopic bodies to be elementary today.
Later, events were redefined by their radar coordinates but
this concept is still classical, since
 an electromagnetic wave is actually an unanalyzed  photon beam.
   There seems to be no experimental evidence
  for an elementary process that
defines only a space-time location,
as general relativity assumes.
In standard theories, nevertheless,
 fermion  events
are  represented
against a classical Minkowski space-time background,
by a $\psi$ vector in an infinite-dimensional complex  Hilbert space.
Both the Minkowski and Hilbert spaces, having  singular groups,
 must be singular contractions
of regular structures closer to experiment.
   An operational  quantum construct of event is still needed.

A single photon is more elementary
 than a radar signal.
The most elementary
quantum events to which we have experimental access today
 are elementary  fermion input/outtake operations.
Let us take these as the elementary events of the next physics, and regard
 the various classical event constructs  as contracted or truncated descriptions.
 
 These more
  physical events have coordinates
  besides position in space-time,
forming a non-commutative algebra.
According to the Standard Model,
 fermionic event today carries  one hypercharge variable $y$,                                                                                                                                                                                                                                                                                                                                                                                                                                                                                                                                                                                                                                                                                                                                                                                                                                                                                                                                                                                                                                                                                                                                                                                                                                                                                                                                                                                                                                                                                                                                                                                                                                                                                                                                                                                                                                                                                                                                                                                                                                                                                                                                                                                                                                                                                                                                                                                                                                                                                                                                                                                                                                                                                                                                                                                                                                                                                                                                                                                                                                                                                                                                                                                                                                                                                                                                                                                                                                                                                                                                                                                                                                                                                                                                                                                                                                                                                                                                                                                                                                                                                                                                                                                                                                                                                                                                                                                                                                                                                                                                                                                                                                                                                                                                                                                                                                                                                                                                                                                                                                                                                                                                                                                                                                                                                                                                                                                                                                                                                                                                                                                                                                                                                                                                                                                                                                                                                                                                                                                                                                                                                                                                                                                                                                                                                                                                                                                                                                                                                    
 three isospin variables $\tau^k$,   
 four space-time position variables $x^m$,
 four momentum-energy variables $p_m$,  
 four spin variables $\gamma^m$\/, 
  eight color charges $\chi^c$, 
  a ternary  generation index $g$,
  and a binary variable distinguishing
  input from outtake operations.
    Let us tentatively omit the generation $g$
  for reasons given in \S \ref{S:QGAUGE}.
 
A simplicial quantum theory
represents a fermion  event by a 
monadic in a subalgebra $\1S[E]\subset \1S$
of some finite type $E$;
call this the {\em event type}.
The first-grade event subalgebra $\1S^{(1)} [E]$ replaces Hilbert space
 as $\psi$
 space to support the algebra $\Alg[E]$
 of single-event operators.
 Higher grades represent composites of several fermions.
The fermion field operators of the Standard Model
form a singular Clifford algebra.
Those of a simplicial theory
are regular Clifford monadics that operate on $\1S[E]$.
They  are elementary in that 
 they are not products of non-trivial factors,
but they generally have a great many ancestral elements of lower type, since
experimental events have high orbital quantum numbers.

The differentiators $\partial_m$ of canonical quantum theory
refer to two levels:  
An infinite-dimensional orbital level of the variables $x^m, p_m $
and a four-dimensional tangent-space level with the index $m$.
The group $\Diff$ acts on the field  and respects the tangent spaces.
Both levels are customarily built into the kinematics by restricting the system to 
differentiable fields and their limits.

In a corresponding class of simplicial quantum theories, monadics $\81_c$ of
a lower type $C$ are used as annihilation/creation operators for the foundational
elements of which events are composed, and
the group of the simplicial quantum theory
is required to respect the partition of the system into cells.

This restriction could be a consequence of the dynamics,
not the kinematics.
Non-differen\-tiable or cell-breaking 
transformations may merely require too much energy. 
The cell level $C$ might be determined by the action function
of the event level, and
the kinematics could have the larger group $\spin(\1S[E])$
of the event space.
This amounts to assuming that the foundation $\psi$ space is 
one-dimensional,
and should be kept in mind.

Each $g$-adic in $\1S^{(g)}[E]$ is  a $\psi$ for a complex of $g$ events
 with Fermi statistics.
The singular Lie algebra of Standard Model  fermion variables
 is to be a contraction of a semisimple subalgebra of
simplicial quantum event  operators in $\Alg[E]$\/.

\subsection{Bipolar quantum spaces}
\label{S:BIPOLAR}

Dirac 
compared a negative probability to a bank overdraft \cite{DIRAC1974}.
Let us expand this somewhat terse  theory of the indefinite metric.

Two frames may share common light cones,
and each may draw its future time axis as a vertical  arrow,
but  may still disagree on the distinction between
past and future,
so that a space-like
 ray $r$ that appears to begin at the origin  in one frame
may appear to end there in another:
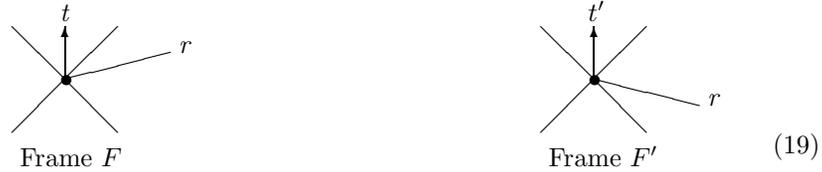
\BEQ
\begin{picture}(200,60)(0,-15)
\put(0,10){$\bullet$}
\put(2,13){\line(4,1){40}}
\put(45,23){\em r}
\put(-15,-20){\rm Frame $F$}
\put(2,13){\line(1,1){20}}
\put(2,13){\line(1,-1){20}}
\put(2,13){\line(-1,1){20}}
\put(2,13){\line(-1,-1){20}}
\put(2,13){\vector(0,1){20}}
\put(0,35){\em t}
\put(200,10){$\bullet$}
\put(202,13){\line(4,-1){40}}
\put(245,3){\em r}
\put(185,-20){\rm Frame $F'$}
\put(202,13){\line(1,1){20}}
\put(202,13){\line(1,-1){20}}
\put(202,13){\line(-1,1){20}}
\put(202,13){\line(-1,-1){20}}
\put(202,13){\vector(0,1){20}}
\put(200,35){$ t'$}
\end{picture}
\EEQ
If the space-time metric is a dynamical variable,
then we cannot restrict our attention to  timelike vectors,
and  an output from the origin can become an input to the origin under this change of frame.
Let us take this to indicate that
when differential locality breaks down,
the distinction between input and output may  become relative,
like the distinction between past and future.

In general,  an operation may introduce an excitation in one circumstance and 
eliminate one in another.
The distinction between input and outtake is then relative to the choice of medium
or vacuum.

Let the difference of two normed spaces
designate their direct sum provided with the difference norm:
\BEQ
 \forall Q\in \1S, \;\forall Q'\in \1S'\;:\quad \|Q\ominus Q'\|:=\|Q\|-\|Q'\|\/.
\EEQ

Minkoski vectors belong
neither to the  space $\1S$,
 nor to $\Dual 3Q$,
 but to their  {\em bipolar} direct sum
\BEQ
\Bi \1S: = \1S \ominus  \Dual \1S\/.
\EEQ

Each experimental frame now splits the bipolar space $\1S[E]$
into two  uni\-polar subspaces of positive-definite and negative-definite norms
respectively.
Let us take flux directed from the experimenter into the experiment
to be positive.
Kets have positive norms and bras have 
negative norms.

The spinor norm thus gives probability flux rather than net probability.
The Lorentz group has regular finite-dimensional  isometric representations
 in the orthogonal group of  $\Bi \1S$.
 
 The ambient spin complex is not random but highly organized locally,
 into something like a crystal dome whose cell is (say) $\SO(3,3)$ invariant.
The dome also
supports the particle spectrum, sharp  bands of highly coherent transmission.

Each  simple Lie algebra of a physical theory has a Killing form
with important physical meaning.
In the original Yang space, it defines a Minkowski metric 
on the energy-momentum vector,
giving the squared rest-energy.
In this way the metrical structure of these quantum space-times
derives from their quantum structure.

\subsection{Simple quantum theories}
\label{S:SIMPLE}

Nowadays  a physical theory involves several
 Lie or near-Lie algebras, some graded.
Let us call a  system  simple or semisimple
if all  these algebras are simple or semisimple, respectively,
and otherwise
{\em compound}\/;
one bad algebra spoils the barrel.
 Since  the commutative coordinate algebra of a classical system
 is compound,
 only a quantum system can be simple.
 Since neither $\slin(\infty)$ nor $\su(\infty)$ is a Lie algebra, let alone a simple one,
 only a quantum system with a finite-dimensional ket space can be simple.
 
 To recover the singular canonical quantum theory (\ref{E:CCR})
  from a simple one like (\ref{E:ROTATOR}),
 one must not only
take a singular commutative limit $h, \4X\to 0$ 
 but must also
freeze the degrees of freedom in the quantized imaginary variable $r=[p,q]$,
 so that $r$ can be treated as a constant $\4N i$. 
This freezing can result from a self-organization of many plexons, like that of 
the spins of a ferromagnet.
 Let us refer to such a construction briefly as a {\em singular organized limit}\/,
and write, for example,
$\5q\cto iq$, $\5p\cto ip$, $\5r/4N \cto  i$.
The symbol $\8 x$ designates a real decontraction 
of a canonical (classical  or  quantum) construct  $x\cfro \8x$.
The symbol $\5 x$ indicates a real decontraction of an imaginary 
canonical construct $ix\cfro \5x$.

The lost  Hilbert space, canonical commutation relations, and Bose statistics must
return in a singular organized limit.

\subsection{Evolution of simplicial quantum theory}
\label{S:EVOLUTION}

Heisenberg effectively introduced a quantum space
 for phase space in 1924, and suggested quantizing space-time in 1930
 as a way to eliminate the remaining infinities.
Julian Wess describes 
how this idea passed from Heisenberg to Peierls, to Pauli, to Oppenheimer,
to Snyder
 \cite{SNYDER1947}, to
C. N. Yang  \cite{YANG1947}, who
made simplified the Lie algebra of the orbital variables $x^{\mu}, p_{\mu}, L_{\mu'\mu}$ 
and $i$
 but still
represented it in the singular algebra $\su(\infty)$ of Hilbert space.

Feynman considered the quantum space (\ref{E:FEYNMAN}) in about 1941,
before undertaking the Lamb shift.

Independently
R. Penrose \cite{PENROSE1971}
quantized the Euclidean 2-sphere
by representing its points as the directions of the sums of many  Pauli spins
with Bose-Einstein statistics.

On still another track to quantum space-time,
Segal \cite{SEGAL1951} made a suggestion already implicit in Yang's note:
that
physics evolves toward simple Lie algebras
by homotopies that reduce Lie-algebra radicals.
 Segal's argument was Darwinian:  
A compound Lie algebra has a singular Killing form, making it labile.
A semisimple Lie algebra is stable in this respect.
 As measurements of the structure constants improve, therefore,
a compound Lie algebra  has survival probability 0
relative to its semisimple neighbors,
which outnumber it by $\infty$ to 1.
Segal wrote this before the Golden Age of Gauge,
but it applies to gauge groups as well.
Singular Lie algebras are indeed  singular cases; almost all Lie algebras
are semisimple.

This Darwinian argument for stabilization must be used with discretion.
For example, 
within the universe of linear spaces with bilinear products
the Jacobi identity is as unstable as commutativity, and it is not proposed here
to stabilize it.
In addition, evolution by small random changes is slow.
Simple organizations not only compete but also cooperate,
and so  form  complex organizations faster than random changes can.
The fact that  relativity and quantum theory
are de-radicalizations is sufficient reason to try another.

Gerstenhaber, influenced by Segal,
described homologically a rich terrain of
 Lie algebras connected 
by  homotopies, such as {\em contractions}
\cite{INONU1952},
 that
carry
groups 
out of  stable valleys of  simplicity,
to  ridges between the valleys, and up to singular
 peaks \cite{GERSTENHABER1964}.
 
 According to the simplicity principle,
physics is flowing glacially down
the simplicity gradient  to a valley in  Gerstenhaberland.
The Galileo Lie algebra, for example,  is on a ridge between the 
valleys of   Lorentz $\so(3,1)$ and the orthogonal group $\so(4)$.
Similarly, the Poincar\'e-canonical Lie algebra $\alg(x^m,p_m, L_{m'm},i)$
is on a ridge between $\so(3,3)$ and $\so(5,1)$.

Group contraction
and deformation quantization were also introduced 
by others \cite{INONU1952,BAYEN1977}.

Vilela Mendez,
 inspired by Gerstenhaber,
rediscovered the Yang $\SO(5,1)$ group \cite{VILELA1994},
representing it in Hilbert space,
and proposed high-energy physical consequences.
The Yang group was rediscovered several times since then.

Galiautdinov  
\cite{GALIAUTDINOV2002},
 Shiri-Garakani
\cite{SHIRI2005},  
and others have
studied  theories with the Yang
$\SO(5,1)$ invariance in $\1S$.
Baugh  \cite{BAUGH2002} represented the Yang 
$\so(5,1)$  Lie algebra in $\slin(6\1R)$.

Now decontraction algorithms  are being  intensively studied \cite{SALOM2011}.

\section{Simplicial quantum relativity}
\label{S:QRELATIVITY}

Assume that the statistics Lie algebra of simplicial events
and  the  Lie algebra of the $\psi$ space are simple
\cite{PALEV1977}.
Represent $\spin(3,3)$ in $\spin(N,N)$,
whose indefinite metric leads to
a {\em bipolar relativity} (\S\ref{S:BIPOLAR}).

Compare the quantized orbital variables of the simple spaces mentioned,
with  $ch\4P\4X$ units and dimensionless orbital variables.
$\8x$  is a quantized $x$, $\5x$ is a quantized $ix$,
$\delta \8q$ is the contribution of one cell to a sum $\8q=\sum\delta \8q$
over many; and
$1\le k\le 3$,  $1\le m\le 4$:
 \BEQ
 \label{E:QCELL}
 \begin{array}{lc|l|l}
\rule[-6pt]{0pt}{18pt}\0{Feynman} & \cite{FEYNMAN1941}& \delta \8x^m \sim \gamma^m&-\\
\hline
\rule[-6pt]{0pt}{18pt}\0{Yang} &\cite{YANG1947}&{\8x^m}\sim i(g^{mn}\eta^5\partial_n-\eta^m\partial_5)
& {\8p_m}\sim i(\eta_6\partial_m-\eta_m\partial_6)\\
\hline
\rule[-6pt]{0pt}{18pt}\0{Penrose} &\cite{PENROSE1971}&  \delta \8x^k\sim\sigma^k&- \\
\hline
\rule[-6pt]{0pt}{18pt}\0{Simplicial}&-&  \delta \5x^m\sim \gamma^{4321}\gamma^{m}&
\delta \5 p_m\sim \gamma_{m}
\end{array}
\EEQ
The proposals of Penrose and Feynman
maintain an absolute distinction between space-time and energy-momentum.
For group simplicity Yang relativized this distinction
 within a relativity Lie algebra
$\so(5,1)$,
but represented this within the singular Lie algebra $\su(\infty)$ of a singular Hilbert space 
$L^{(2)}(6\1R,\1C)$.
This respects both reciprocity and parity.
For the sake of regularity, 
the simplicial quantum theory represents both time and energy
within  a spin representation of  Lorentz $\spin(3,1)$ instead.
This also introduces fewer variables.
Furthermore it
violates parity and reciprocity,
in better accord with experiment.
The factor $\gamma^{4321}$ has been shifted from $\delta p$ to $\delta x$ 
to make the gaugeon dynamics easier.

Unitary charges of the Standard Model kind
are short transverse struts to the dome.
The cellular plexons of the ambient dome 
therefore have a higher dimension than classical space-time,
are spins of a higher orthogonal group.
Dirac spins are short longitudinal struts.
Space-time coordinate lines are longitudinal and long.
If they include transverse struts,  their unitary charges cancel within a few cells.
The quantized imaginary $\Qi$---where $\supo{}$ stands for ``operator"---is macroscopic  and composed
of longitudinal struts, and so can be classed as longitudinal,
but it is polarized and frozen.
There are now three non-zero vacuum expectation values to account for:
\begin{itemize}
\item [g.] Gravity's  $g_{\mu\nu}$, which  breaks space-time $\slin(4\1R)$.
\item [h.] Higgs $h_{i}$, which breaks electroweak $\su(2)$.
\item [i.] The $i$ in the Heisenberg equation and the canonical commutation relations,
which
breaks complex-plane $\slin(2\1R)$ and time reversal $\1Z_2\subset \SL(2\1R)$.
\end{itemize}

Let us represent the dome and its excitations within $\1S$
using as few $\iota$ types as possible.

When $i$ becomes a variable $\Qi$ and leaves the center of the algebra,
 a real quantum theory arises
of the St\"uckelberg kind
 \cite{STUECKELBERG1960}.
This also recalls the Hestenes theory in which  $i=\gamma^{4321}$
 \cite{HESTENES1966},
and the quaternionic $i$ that was proposed as a mass-generating Higgs field
\cite{TAVEL1965}.

The canonical relations work
in a segment of the  spectrum of $-\Qi^2$ so near to its maximum value 1
as to be indistinguishable from it in atomic experiments.
Yet this narrow band must have a multiplicity that passes today for infinite.
For but one example, the band
\BEQ
1 - {\4N}^{-1/2}<  -\Qi^2 \le  1\/
\EEQ
is both narrow and populous enough, with
 multiplicity $\0O( \sqrt N)\to \infty$.

In the singular limit ${\Qi}\to i$, $E\to 0$,  and (supposedly) 
$(\Qi)^2\to -1$,
 the classical space-time  and  canonical commutation relations are to emerge.  
We must suppose that
\BEQ
 \4X\4P\4N=\hbar,  \quad \4N \gg 1, \quad h/\4X\gg 1\, \mbox{TeV},\quad  
 \frac h{\4N\4X}\approx 0,
 \EEQ
in the sense that $h/\4N\4X$ is presently  not resolvable from 0.

To achieve this one may
hypothesize an organization akin to polarization that centralizes (``superselects") the
variable $\Qi$ and contracts it to
the imaginary unit of complex quantum theory:
$\Qi\cto i$\/.
The Heisenberg relation $[x^m,p_m]=\hbar i$ returns in a singular limit
of many cells.
This means zero-point estimates based on canonical quantum theory
are gross overestimates in the cell domain,
including the ones that call for infinite renormalization and energy density.

The coordinates, momenta, and $\Qi$ are now
cumulated spin variables.
To make all components
of a  cumulated spin small is not impossible
in the way that making  position and momentum coordinates small is supposed to be.
It may merely require a meltdown of $\Qi$.

\subsection{Reciprocity and locality}
\label{S:RECIPROCITY}

In canonical classical and quantum theories and all their    regularizations considered here, the kinematics obeys reciprocity
while experience seems to flout it,
especially by its locality.
The locality
principle
permits fields to couple only when they are evaluated at the same position,
but different momenta (wave numbers) generally couple.

In the simplicial quantum space (\ref{E:QCELL}),
reciprocity is the 
transformation 
\BEQ
R:\gamma^m\mapsto \gamma^{4321}\gamma^m
\EEQ
$R$ is  an automorphism of the Dirac Clifford algebra.
The simplicial event space has reciprocity.
Locality presently breaks it.
In the simplicial quantum theory, however,
a simplicial ``ultra-locality" is  tautological:
Two simplices $\gamma^A, \gamma^B$ grade-commute unless they are neighbors,
share a vertex $\gamma^a$, $a\in A, B$.

\subsection{Palev statistics}
\label{S:SPINSTATISTICS}

Bose-Einstein statistics is a defining though singular part of canonical quantization.
Fermi-Dirac statistics  generalizes from Lie algebras
 to  graded Lie algebras, 
such as Grassmann algebras,
in order to deal with variables without classical correspondents,
like  fermion creators and spins.

A {\em Palev statistics} (of class $\4a$) is one whose one-quantum 
annihilation/creation operators generate
a semisimple graded Lie algebra $\4a$. 
Fermion pairs are not exactly bosons,
only quasi-bosons,
but they are exact Palevons.
They can simulate bosons as
polarized  spins can simulate an oscillator
by precessing about the polarization axis.

Most attention has been given to the simple ungraded case;
but Stoilov and Van der Jeugt find that the
Palev statistics of class $\su(1|5)$ is especially relevant to the Standard Model
\cite{STOILOV2005}.

In pure Grassmann simplicial quantum theories, like the present study,
one must represent   empirical bosons  as event dyads.
These obey
a Palev statistics,
with fermionic cores.
Presumably this fermion core must show up in high-energy photon-photon collisions.
Arguments against composite photons or gravitons are well known,
but are based on Bose-Einstein statistics and the canonical commutation relations,
and therefore must break down in exactly the domain of interest.

With this generalized concept of statistics,
 any quantum number can be regarded as a number of quanta.
This is familiar for the linear harmonic oscillator,
where the quantum is
the phonon of excitation.

For instance,  the quanta of the dipole rotator in three dimensions,
with Lie algebra  (\ref{E:ROTATOR})--call them rotatons,
since the term ``roton" is preempted---have a statistics that is
neither
Bose-Einstein nor Fermi-Dirac,
but Palev  of the $\spin(3)=A_1=B_1$ class.
The individual kinds of Palevons correspond to
root vectors of the Palevon Lie algebra.
The statistics of the Palevon is determined by an irreducible representation
of the Palev Lie algebra.
Where the canonical commutation relations have just one
practical representation,
each Palev statistics has infinitely many.
The $\so(3)$ Palev statistics is isomorphic to the 
rotator (\ref{E:ROTATOR}), and
has only one kind of quantum and its dual.
The Bose-Einstein statistics 
arises from the
representation  $D(N)$ of 
(\ref{E:ROTATOR}) 
when $N$ is allowed to approach infinity
with organization that polarizes and centralizes
one component, say $r$.
The corresponding physical theory, however,
 has a finite $N$
that must be measured.

The same $\so(3)$ class of Palev statistics
includes the spinor representation $D(1/2)$.
Its representation in $\1S$ requires
spinors of four real components and is then 
given by
\BEQ
p=\gamma^{32}/2,\quad q=\gamma^{21}/2,\quad r=\gamma^{13}/2.
\EEQ
A spin $1/2$ is thus a composite of up to 2
rotatons with Palev statistics.
It is necessary to see how Palev statistics describes
a condensation resembling the Bose-Einstein one.
This requires many Palevons, $N\gg 1$, in at least two classes,
one to be filled and the other to be emptied.
 
 A monadic of any type
embraces monadics of the type below.
As a result, lower-type operators $L[T]$
induce higher-type ones $L[T+1]=\Sigma L[T]$.
In particular,
quantum spin operators 
of the ancestral cell induce  orbital angular momentum operators
of the event.
The cumulant $\Sigma^n L[T]$  represents the operator $L\in \Alg(\1S[T])$ in 
the algebra $\Alg(\1S{[ T +n]})$ of every  higher type $T+n$.

Let us apply the cumulation process to the 
Lie algebra of the simplicial quantum space-time of (\ref{E:QCELL}).
Let $\gamma^{\ddot b}$ be four real Dirac spin operators acting on
spinors $1_c$ of a cellular level $\1S[C]$, with Minkowskian metric form 
\BEQ
\label{E:DIRAC}
g^{{\ddot b}'{\ddot b}}=
\frac 12 [\gamma^{{\ddot b'}}, \gamma^{\ddot b}]\/.
\EEQ
The ancestral  group generators $\delta J_{c'c}[C]$ of the cellular spin are
\BEA
\label{E:QSTEM}
\begin{array}{lll}
\delta \5 x^{\ddot b} &=& \4X\,\gamma^{{\ddot b}},\quad {\ddot b}=1,2,3,4 \/,\cr
\delta\5 p_{\ddot b}& =& \4P \,\,\gamma^{4321}\gamma_{{\ddot b}}\/,\cr
\delta\,\Qi &= &\4N^{-1}\, \gamma^{4321}\/ , \cr
\delta\5L_{{\ddot b}'{\ddot b}}&=&h \gamma^{4321}\gamma_{{\ddot b}'{\ddot b}}\/,\quad \mbox{as}\cr
\0h(4)& \cfro  &  \so(3,3)\;\cfro \;\slin(6)
\end{array}          
\EEA
The simplicially quantized imaginary $\Qi$ is 
normalized to unit magnitude with a factor $\4N^{-1}$.
To form macroscopic monad coordinates, 
we  must cumulate these atomic cell variables at least twice,
to
reach at least
type 6, with dimensionality 
 $\hexp 6=2^{64\0K}$, more than ample for a quasi-continuum.
 
Consider  six $8\x 8$ $\gamma^{\ddot b}$ of the cellular type $C$ 
associated with $\spin(3,3)$.
Choose a frame 
with metric $\ddot b=\mbox{diag}(1,1,1,-1,-1,-1)$
and label the $\gamma$'s as  $\gamma^1,\dots,\gamma^4,\gamma^{X},
\gamma^Y$, with
spontaneous polarization of $\gamma^{YX}$:
$\Qi=\Sigma^2 \gamma^{YX}[C]\cto i$\/.
Define the  Dirac $\gamma^m$ as the remaining four first-grade elements of $\gamma^c[C]$.
They define the Minkowski space-time by cumulation and the singular limit.

\subsection{The Umklapp problem}

When Heisenberg proposed to quantize space-time,
Pauli  pointed out the Umklapp problem \cite{WESS2005}.
Feynman raised it for his quantized space-time.
Discreteness of the  space coordinates with spacing $\4X$, 
\BEQ x=n\4X, \quad
(n\in \1Z),
\EEQ
ordinarily
implies periodicity of the  momenta with period $2\pi \4P=2\pi \hbar/\4X$:
\BEQ
p\cong p+2\pi n\4P.
\EEQ
Then the maximum momentum $p=\pi n \4P$  in one direction
is indistinguishable from 
the maximum momentum $-p$ in the reverse direction.
Phonons  in crystals flip their momenta  by this process,
where the missing momentum is taken up by the crystal;
but not photons in the vacuum.

Simplicial quantization creates no Umklapp problem.
The spectrum of the  momentum $\4P L_{mY}$  is still bounded by $\pm \4N\4P$, 
and  is still discrete and uniformly spaced between these bounds,
as in a crystal,
but it is not uniform in spectral multiplicity,
and  $L\doteq \4N$ is not the same eigenvector as $L\doteq-\4N$.
None of the variables $L_{c'c}$ is periodic.

Furthermore,
to attain a maximum value for $L_{mY}$
requires  polarizing all $\4N$ spins in the sum along  the $mY$ direction.
Since $L_{mY}$ does not commute with $L_{YX}$,
this disrupts the dome polarization of the $L_{XY}$  that produces $\Qi$,
and so must be difficult, perhaps unfeasible.

\section{Simplicial quantum gauge}
\label{S:QGAUGE}

Gauge fields are associated with non-commutative fermion momentum variables.
A canonical gauge  field theory with  a semisimple gauge Lie algebra $\4g$
splits the total momentum-energy $p_m$  into an invariant kinetic part  
$\pi_m$
and a residual potential part $\Gamma_m$:
\BEQ
p_m=\pi_m +\4P \Gamma_m
\EEQ
$p_m$  is integrable ($[p_m,p_{m'}]=0$) but not gauge invariant.
$\pi_m$ is gauge invariant but not integrable.
$\4P$ is a constant with dimensions of energy-momentum ($c=1$).

Simplicial quantization also produces non-commutative momentum components.
Since the ancestral atoms of momentum do not commute,
 neither do their cumulants, the simplicial quantum momenta, the reformations of
 the singular infinitesimal translations
 canonical quantum theories.
Some of this quantum non-commutativity survives into general relativity as part of the curvature,
perhaps including a cosmological constant contribution from the ground complex, 
a gravitational   energy-momentum (``dark energy'').
We can be certain that this contribution is finite,
and reasonably sure that it is much smaller 
than the canonical commutation relations permit,
since the mean magnitude of the zero-point energy of a simplicial quantum field oscillator
is less than its canonical correspondent by a cosmologically large  numerical factor.

Therefore it is natural to ask
 whether gauge fields are simplicial quantum effects expressed
 in the canonical quantum limit; much as Poisson Brackets are 
canonical quantum effects expressed in the classical limit.
 Let us call such a gauge theory simplicial.
 
 In a more canonical theory, the gauge vector fields are ``internal" parts of the 
 gravitational field \cite{CONNES1994}.
 In a simplicial theory, the gravitational field is an external self-organization
 of many plexons.

In the simplicial quantum gauge theory,
 the momentum-energy $\5p$, positional coordinates $\5x$,
 and quantized imaginary $\Qi$
are generators of $\slin(6)\sim \spin(3,3)$,
 represented as polyadic fermion processes.

Since gaugeons do not obey Fermi statistics they cannot have monadic $\psi$'s.
The simplest possibility is dyadic, as if they were fermion pairs.
Dyadic plexons obey a Palev statistics,
and the gauge group is their Palev Lie algebra.

De Broglie suggested that photons are neutrino pairs,
a special case.
In the canonical quantum theory this seems untenable.
According to the Heisenberg uncertainty relations,
when two identical fermions are close in their position coordinates they must be
far apart in their momenta on the average.
The forces binding them would have to be large for high relative momenta.
There seem to be no such forces.
On the contrary, asymptotic freedom seems to be observed, as well as 
inferred by Gross and Wilczek from non-abelian quantum gauge theory.

The simplicial quantum gauge theory reopens this classic question 
by weakening the 
Heisenberg uncertainty principle enormously, allowing
small values of both relative position and relative momentum
at once when the organization of $\Qi$  is highly broken.
In the case of a definite gauge metric and an irreducible group,
the conservation of the Casimir operator still blocks
the possibility that $p,q,r$ can all be small.
But the simplicial quantum metric is indefinite;
its Casimir operator can be constant even when its terms
decrease.

There are other indications for dyadic simplicial  gaugeons:

\unit{\em Spin.}
The spins and the statistics add up correctly.  
This is what motivated De Broglie
to propose the fermion-pair theory of the photon and 
Feynman to consider a fermion-quartet for the graviton.
The graviton is the noblest particle,
in the way that helium is the noblest gas
and the alpha particle a noblest (that is, most magical) nucleus.
It would be natural to suppose that, like them,
it is composed of  four fermions that have saturated each other's valences.
Here it is supposed that it and the photon are composed of two
fermions that have saturated each other's valences.

The Minkowski index $m$ of the gaugeon vector $\Gamma_{m\alpha}(x)$,
where $\alpha$ indexes a basis for the gauge Lie algebra,
contributes  $\hbar$  to the gaugeon spin angular momentum, as usual.
The two fermions in the dyad
are  at slightly different quantum positions
and can orbit about each other.
The index $m$ indicates a relative orbital angular momentum of $1\hbar$.
 
The Lie algebra index $\alpha$ contributes another unit spin
 for the graviton, where the unitary spins are saturated and $\alpha$
 indexes the
 gauge Lie algebra $\spin(3,1)$;
and  not for the unitary gaugeons, where the fermion spin valences are saturated.
Thus all the gaugeons, including the graviton, can be fermion-dual-fermion pairs,
plus higher-order terms of 
more complex structure.

Every fermion has spin 1/2 and Fermi-Dirac statistics, and some have
 color or isospin.
 It is well-known how to represent fermions in Hilbert space.
 Then the fermion basis $\psi$'s are tensor products of  more elementary $\psi$'s
supporting orbital variables, spin, isospin, color, and generation number.
These are its valences.
Since they are distinguishable, there is no physical change
if Grassmann products are used instead of tensor products.
Then the whole construction is readily carried out in $\1S$;
this adds no insight as yet.

As for the gaugeons: The generic gaugeon has as semisimple gauge group
the semi-simple product of all the simple gauge groups of the fermions.
If it is a pair composed of just one kind of fermion,
the fermion must have everything,
so it is a quark.
The isospin gaugeons are pairs of quarks in which all the
color valences are saturated;
the color gaugeons saturate their isospin valences;
gravitons saturate
both, leaving only their spins open.

The pair theory of gaugeons has a generational problem, however.
There are three generations $g_1, g_2, g_3$ of fermions,
and nine possible pairs   $(g_ig_j)$,
but only one generation of gaugeons is seen.

In one resolution of this conflict,
one quark generation  $g_1$ would be more elementary than the other two,
and gaugeons are  pairs $g_1g_1$ of quarks of that 
generation.
The other two generations  $q_2, g_3$ of fermions may then be
block from pairing by their structures.
In that case, however,they would couple 
 to  gaugeons more weakly than  the  generation $g_1$.
This would disagree with the Standard Model.

In a resolution that saves the Standard Model,
only the symmetric superposition $g_1g_1+g_2g_2+g_3g_3$ is a stable gaugeon.
This gaugeon couples to all generations alike.

\unit{\em Connection.} A fermion dyad $\psi\v \5{\psi}$ 
(where the tilde indicates the dual particle)
 indeed transports a third fermion from $\psi'$ 
to $\psi$ as a gauge connection should, according to
\BEQ
\psi''=(\psi\v \5\psi')\circ \psi'\/.
\EEQ

\unit{\em Coupling} If the gaugeon is a fermion pair,
the Dirac action of the canonical gauge theory musy be a contraction of a polyadic simplicial action.
Let us de-contract it:

\subsection{Simplicial gauge dynamics}
The Dirac fermion-gaugeon coupling density $A_{\0D}$ is  an invariant trilinear form
 in annihilation/creation operators 
$\psi(x) $ for a fermion,  $D_m(x)$ for a gaugeon, and  $ \6{\psi}$
for a dual fermion:
\BEQ
\label{E:DIRACTION}
A_{\0D}(x)=i\6\psi(x) \gamma^m(x) D_m(x)\psi(x) = {>}\!\!{\bullet}\!{=},
\EEQ
in which ``$>$'' designates the two fermions and ``$=$'' the gaugeon.
All four factors are spin operators with invisible spinor indices,
and are multiplied accordingly.
Schematically speaking, $A_{\0{D}}$ couples a 
vector $\6\psi\gamma^m\psi$ of the fermion pair to a vector
$D_m$ of the gaugeon at infinitesimally separated space-time points;
it is a V-V coupling.  
Clearly one can form S-S, V-V, T-T, A-A, P-P couplings,
and S-P and V-A couplings,
 as in early theories
of beta decay.

The system history $\psi$ is the time-ordered exponential
\BEQ
\Omega = \Texp \int(dx)A_{\0D}(x)\/.
\EEQ
$A_{\0D}$  is required to be skew-hermitian so that $\Omega$ is  unitary.

(\ref{E:DIRAC}) defines a Clifford algebra only when the gravitational metric 
$g^{m'm}(x)$ is treated as a constant quadratic form.
When quantum gravity comes into the action, this space-time Clifford algebra breaks down.
On the other hand, there is no reason to assume that the Fermi-Dirac Clifford algebra
also breaks down, so
it is still reasonable to represent
quantum gravitational processes with spinors in $\1S$.

Let us infer from the theories of the other gaugeons that
the simplicial gravitational gaugeon too is described
by a simplicial correspondent  $\8{D}$ of the differentiator $D_m$, 
with curvature field $[\8D_{m'}, \8D_m]=\8R_{m'm}$;
not (say) by the simplicial correspondent of the Dirac spin vector  ${g^m}$, 
since (\ref{E:DIRAC}) breaks down.

Suppose that the gaugeons in $D_m$ are bound fermionic pairs,
associated with two proximate events rather than one,
as discussed in \S \ref{S:SPINSTATISTICS} and \S\ref{S:QGAUGE}.
Then the Dirac action (\ref{E:DIRACTION}), 
which couples two fermions and a gaugeon,
is actually a surrogate action for 
a four-fermion (``Fermi") coupling, effective when two of the fermions are bound;
a tetradic
\BEQ    
\label{E:TETRADIC}     
A^{(4)}
 \sim \Gamma ^{e'''}{}_{e''}{}^{e'}{}_{e}  \;   \6\psi_{e'''}\6\psi^{e''}{\6\psi}_{e'}\6\psi^{e}
\EEQ
in which  $(\Gamma)$ is an invariant coupling tensor with real components.

For a physical theory with this tetradic coupling to be possible,
the time development should be an orthogonal  operator and its generator
 $A^{(4)}$ must be skew-symmetric with respect to the spinor metric form $\beta$: 
\BEQ
(\beta A^{(4)})^{\4T}=-\beta A^{(4)}\/,
\EEQ
where  $(\dots)^{\4T}$ is the transpose of $(\dots)$\/. 
This
holds for  four-dimensional spinors of  $\spin(3,1)$,
whose $\beta$ is skew-symmetric,
 if $(\Gamma ) $ has the symmetry property
 \BEQ
\Gamma ^{e'''}{}_{e''}{}^{e'}{}_{e}  =\Gamma ^{e'}{}_{e}{} ^{e'''}{}_{e''}{}.
\EEQ

The  simplest-looking tetradic action 
\BEQ
A^{\0{SS}}\sim ({\6\psi}\circ\psi)({\6\psi}\circ\psi)
\EEQ
also has this property;   but violates locality grossly, since
every event in one pair is coupled to every event in the other.
In the Dirac action, only infinitesimally separated events are coupled.

The simplicial action closest to the Dirac action is the V-V coupling
\BEQ
A^{\0{VV}}:=(\6\psi \gamma^{\ddot b}\psi)(\6\psi \gamma_{\ddot b}\psi) \in \Cliff \3W[E]
\EEQ
Each factor is a simplicial tensor  with invisible indices, whose 
structure is defined so that it corresponds to that of the Dirac action.
Since $\gamma^{\ddot b}$ puts in or takes out a single vertex,
it couples only adjacent plexons, and this may suffice for the experimental locality.
A more explicit formulation is reserved for the computational stage.

Turn from fermions to gaugeons. 
The usual gaugeon action of the Standard Model has
the form 
\BEQ
A_{\0G}\sim [D_{m'},D_m][D^{m'},D^m]\/.
\EEQ
This becomes an eight-fermion coupling,
effective when all eight fermions are bound into four pairs.
A corresponding plexon action  is octadic:
\BEQ    
\label{E:OCTADIC}     
\8A^{(8)}
 \sim \Gamma ^{e^8}{}_{e^7}{\dots}^{e'}{}_{e}  \;  
  \6\psi_{e^8}\psi^{e^7}\dots{\6\psi}_{e'}\psi^{e}
\EEQ
in which  $(\Gamma)$ is an invariant coupling tensor with real components.

It is impossible to repress the speculation that the octadic gaugeon action 
$A^{(8)}$ is an effective action 
of second order in the tetradic action,
operative when eight fermions bind into four pairs.
In the continuum-based theory such speculation may be idle, due to infinities,
but in the simplicial theory
the coefficients, including the fine-structure constant,
are mathematically well-defined,
and probably computable for sufficiently small-scale experiments.

\subsection{Space-time curvature as  quantum effect}

The classical momentum components $\partial_m$ commute with each other,
but the gauge-covariant momentum components $i\hbar D_m$ do not, 
nor do the simplicial quantum components $\8p_m$.
This must contribute to space-time curvature
and the gauge tensor fields.
It is parsimonious to conjecture that
 the classical non-commutativity of the gauge differentiator,
like that of the Poisson Bracket,
is entirely a vestige of the quantum non-commutativity of the pair annihilation/creation operators.

The  simplicial quantum curvature is the quantum commutator:
\BEA
\label{E:CURVATURE}
K_{{\ddot b}'{\ddot b}}&=&{\4P}^2 [J_{Y{\ddot b}'}, J_{Y{\ddot b}}]\cr
&=& - {\4P}^2 J_{{\ddot b}'{\ddot b}}\/,
\EEA
in a frame where the form $b$ is in Sylvester normal form with $b_{YY}=-1$.
$K$ includes the gravitational curvature $R$ and the  unitary gauge curvatures
as non-unified terms in a sum.
In the simplicial quantum theory the commutators
$[J_{e'''e''}[E], J_{e' e}[E]]$ are related to
other components of the same generating tensor $J$
 by the structure tensor $c$ of $\spin[E]$.
Thus the infinity of differential concomitants of the
gauge manifold, including the curvature, are replaced
in the simplicial quantum theory by the finitely many components of
the generating tensor $J[E]$ of type $E$. 

\section{Discussion}
\label{S:DISCUSS}

Spinors belong to a Grassmann algebra (Cartan);
a Grassmann algebra represents 
a simplicial complex (Chevalley). 
It has long been suspected
 \cite{BOHM1962, MERMIN1971, KLEINERT1989}
that quantum gauge theory
might be a singular limit 
of a  simplicial complex theory, 
and that the gauge charges 
 are vestiges of quantum Burgers(-Volterra) vectors 
 for defects of various kinds.
 Then the manifold-based gauge theory is not fundamental but a
  high-level consequence of the underlying cellular dynamics
 of the complex.
 
 The simplicial quantum theory presented here
represents the quantum system as a simplicial quantum complex
and its
 gaugeons, including the graviton, as
 fermion-dual-fermion  pairs.
Differential transport is  a
 a singular continuum limit 
 of  annihilation reactions of the form
 \BEQ
 (\psi \v{ \psi'}^{\beta} )\v \psi''\to \psi,
 \EEQ
where a pair effects a quantum transport
 by replacing the fermion $\psi''$ outside the pair by 
the fermion $\psi$ inside the pair.
The classical
 principle of gauge invariance is not basic
 but emerges from the  simplicial quantum structure of the
deeper levels as a smoothing approximation.

New quantum constants enter: 
a large integer $\4N$ limiting the number of events,
  and quantum units $\4X$ and $\4E$ of time and energy.
The
usual imaginary $i$ becomes a quantum spin operator,
which has to be frozen and centralized  by a spontaneous organization in the vacuum.
The bras and kets belong to a space $\1S$,
a real Grassmann algebra  over itself, not only graded but also 
typed in the sense of Quine.
Its first-grade elements are spinors.
Its neutral spinor metric assigns  positive norms to kets, negative to bras.

 This paper is a palimpsest.  
 Between its lines can dimly be made out remnants of
an earlier work called the space-time code \cite{FINKELSTEIN1969,FINKELSTEIN1996}.
Instead of quantizing the Standard Model, working from the top down,
the space-time code attempted to quantize space-time from the bottom up.

The weakened uncertainty relations for small distances and momenta
  make it possible that
gaugeons, including gravitons,
 are fermion pairs, as their  transformation properties suggest.

A plausible fermion-gaugeon action  is then a  tetradic  (\ref{E:TETRADIC})
 in fermion annihilation/creation operators, summed over the cells of the complex.
Additional action terms 
 like the Maxwell, Hilbert-Einstein, and Yang-Mills actions
 for the separate gaugeon fields are not necessary
if they are effective actions for organizations
derived from (\ref{E:TETRADIC}).
 Then the gauge coupling constants  are order parameters of this organization.
 
The road from 
dynamics  to particle spectrum and 
cross-sections is
not yet entirely  computational.
Existing quantum concepts of binding and scattering
rest on  unstable classical foundations in Minkowski space-time
and idealizations like stationary or plane  $\psi$ waves.
They too must now be transferred to the
more stable quantum foundations of
the quantum complex, finite  in both the large and the small.
 
 \section{ Acknowledgements and references} 
 
Roger Penrose and  Richard Feynman graciously
 showed me  unpublished spin quantizations of space or space-time that 
were most important for this work.
 I am obliged for other helpful communications
 to  Otto B. Bassler,
James {Baugh}, Walter L. Bloom, Jr., Dustin Burns, David Edwards,
 {Shlomit} Ritz Finkelstein, Andrei {Galiautdinov}, Josef Maria Jauch,
 Dennis Marks,
  Zbigniew Oziewicz, David Speiser, Tchavdar Palev, Heinrich {Saller}, Stephen Selesnick,
  Abraham Sternlieb,  Sarang Shah,
Frank (Tony) Smith, and Carl-Friedrich von Weizs\"acker.

\bibliographystyle{plain}
\end{document}